# Recommending Insurance products by using Users' Sentiments

Rohan Parasrampuria[1], Ayan Ghosh[2], Suchandra Dutta[3] and Dhrubasish Sarkar[4(*)]

[1, 2, 3, 4]Amity Institute of Information Technology, Amity University Kolkata, India
`¹rohanparasrampuria@gmail.com`, `²ayanmailbox187@gmail.com`, `³dut-tasuchandra214@gmail.com`, `⁴dhrubasish@inbox.com`
`(*) Corresponding author`

**Abstract.** In today's tech-savvy world every industry is trying to formulate methods for recommending products by combining several techniques and algorithms to form a pool that would bring forward the most enhanced models for making the predictions. Building on these lines is our paper focused on the application of sentiment analysis for recommendation in the insurance domain. We tried building the following Machine Learning models namely, Logistic Regression, Multinomial Naïve Bayes, and the mighty Random Forest for analyzing the polarity of a given feedback line given by a customer. Then we used this polarity along with other attributes like Age, Gender, Locality, Income, and the list of other products already purchased by our existing customers as input for our recommendation model. Then we matched the polarity score along with the user's profiles and generated the list of insurance products to be recommended in descending order. Despite our model's simplicity and the lack of the key data sets, the results seemed very logical and realistic. So, by developing the model with more enhanced methods and with access to better and true data gathered from an insurance industry may be the sector could be very well benefitted from the amalgamation of sentiment analysis with a recommendation.

## 1 INTRODUCTION

Customer satisfaction which is regarded as the key to success in any business is best achieved by analyzing the needs of the customers and solving the problems that they may be facing when using the products. The customer feedback lines regarding the product seem a very important source of insight as to what the users think about the product and which user group likes the product the most, what can be done to make their experience even better. Mining the sentiments inherent in any customer feedback line can help businesses not only understand better about their customers but also about their products. In this paper, we have chosen insurance as the concerned business and we have tried to recommend products to a customer by analyzing the sentiments from their feedback lines. But, some basic terms first,

## 1.1 Recommendation System

## 1.2 Machine learning

Machine Learning is the science of getting machines to act without being explicitly programmed, it is based on algorithms and a bunch of Artificial Intelligence focused on building applications that learn from data. In data science, an algorithm is a statistical processing step. Whereas Machine Learning algorithms are trained to find patterns and features in big amounts of data to make decisions and predictions based on new data.

## 1.3 Sentiment Analysis

Sentiment Analysis is often termed opinion mining. Sentiment Analysis is normally concerned with the voice of client materials. It is the process of analyzing online pieces of writing to determine the emotional tone they carry, whether they're positive, negative, or neutral.

## 1.4 Random Forest

Random Forest is also called a random decision forest. It's a machine-learning algorithm that relies on ensemble learning for classification. It is a method of ensemble learning.

## 2 Related Work

The suggested technique uses ordinal categorization to determine the polarity of user evaluations [1]. A machine learning comparison review in sentiment analysis showed how a random forest classifier shows results with greater accuracy [2]. Lexicon-based approach, Sentiment Analysis Using Random Forest Ensemble for Mobile Product Reviews in Kanna-da [3]. An ensemble sentiment The Twitter data analysis method for airline services analysis shows the use of naïve Bayes and Bayesian network [4]. User's profile and sentiment analysis are important factors in recommendation system, Music recommendation system based on user's sentiments extracted from social networks [5]. Recommendation models which are based on text re-views and temporal influences, a tourism destination recommender system using users' sentiment and temporal dynamics [6]. Random Forest is an accurate and robust classifier, Random Forest with Tuning Hyperparameters for Sentiment Analysis of Movie Feedback [7]. A Study of the Aspects Affecting the Cost System 's Usability Decision Trees, Support Vector Machines, and Logistic Regression were used to demonstrate how data was obtained to use a question sample, which is contextual and represents the participants' expectations. [8]. A five-point scale, that corresponds to the product ratings occurring in the corporate world. A five-point scale is nothing but ordinal classification, showing the use of Logistic Regression [9]. Classification of Sentiment Analysis on Tweets using Machine Learning Techniques portrays the use of multinomial naïve Bayes and its shortfalls with others [10]. A hybrid mechanism- 'Cluster-then predict Model' improves the accuracy of prediction [11]. Recommender systems (RS) have become a hot topic in the study, intending to assist consumers in finding products online by making choices that closely match their interests [12]. The online social network aims to assist individuals by utilizing the information provided by victims of crime to minimize future criminal activity. People will profit from learning about various criminal actions, as well as who the victims are and how they've been affected [13]. Some approaches, aspects, and drawbacks of information diffusion are described in this study [14]. For any of the defined social fields with sentiment kinds, this article has developed an innovative and efficient influence assessment methodology to estimate an impact factor of each influential interacting node in the user's network [15]. Utilizing vector space models and phrase frequency-inverse document frequency approaches, a statistical model is provided in this paper to estimate the behavior acceptance amongst users in various timestamps on online social networks [16]. By studying and classifying behavior, the model employed subjectivity and polarity to create a tailored recommender system [17]. The goal of this article is to utilize a community detection algorithm to identify a link between criminal events and user profiles, with a vector space model as a crucial component, and then to apply a recommendation algorithm on users to advise the results of the previous study [18].

## 2  Block diagram

Following diagram **Fig.1.** shows the basic structure of our model.

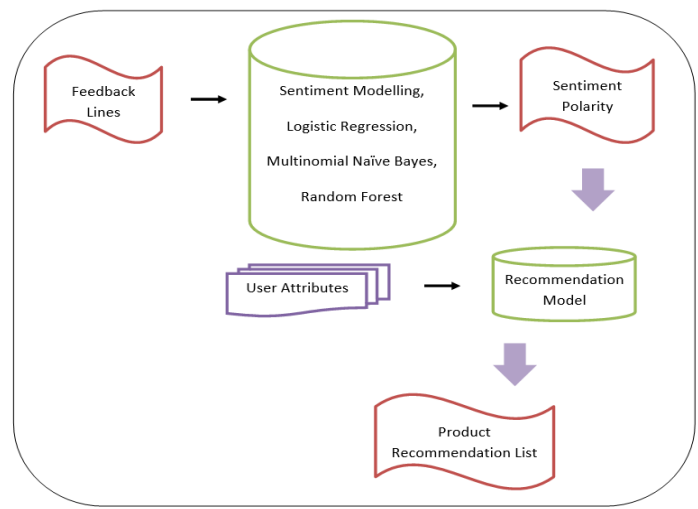

**Fig. 1. The Block diagram of our model**

## 3       Proposed model.

Following model **Fig.2**. will explain the step-by-step process of the whole work.

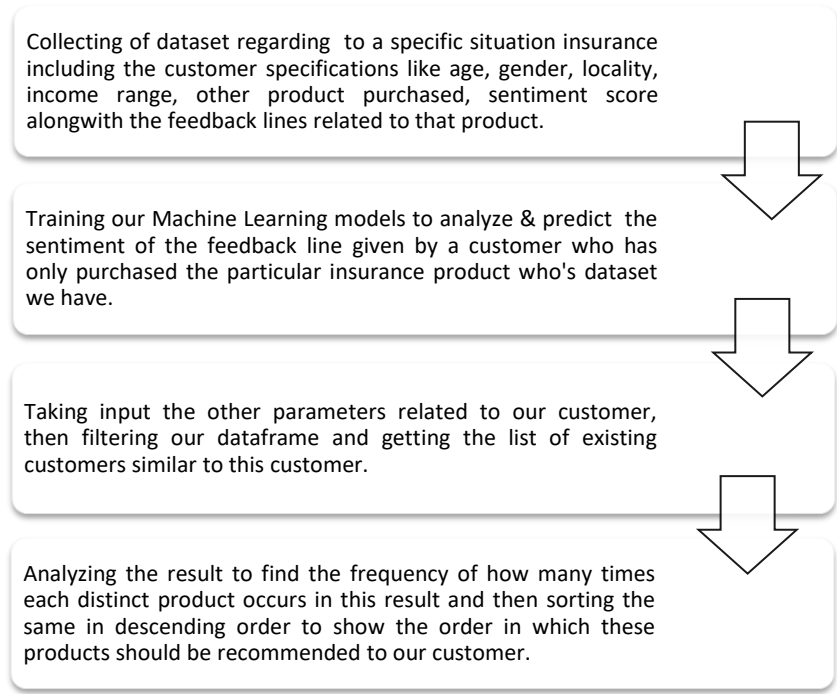

Collecting of dataset regarding to a specific situation insurance including the customer specifications like age, gender, locality, income range, other product purchased, sentiment score alongwith the feedback lines related to that product.

Training our Machine Learning models to analyze & predict the sentiment of the feedback line given by a customer who has only purchased the particular insurance product who's dataset we have.

Taking input the other parameters related to our customer, then filtering our dataframe and getting the list of existing customers similar to this customer.

Analyzing the result to find the frequency of how many times each distinct product occurs in this result and then sorting the same in descending order to show the order in which these products should be recommended to our customer.

**Fig. 2. Shows the step by step working of our model**

## 4       Result Analysis

### 4.1     Sentiment Modelling

   a.   **Logistic Regression**

   Using the Logistic Regression, the model gave the following,
   Accuracy = 87.88%
   False Positive Count = 457
   False Negative Count = 452
   The Confusion Matrix is shown in below picture **Fig.3.** and Accuracy Calculations produced using the Logistic Regression model,

```
Confusion matrix:
[[3287  452]
 [ 457 3304]]
Accuracy:
0.8788
```

Fig. 3. Confusion Matrix and Accuracy Calculations – LR

### b. Multinomial Naïve Bayes

Using the Multinomial Naïve Bayes, the model gave the following,
Accuracy = 84.4%
False Positive Count = 702
False Negative Count = 468
The Confusion Matrix is shown in below picture **Fig.4**. and Accuracy Calculations produced using the Multinomial Naive Bayes model,

```
Confusion matrix:
[[3271  468]
 [ 702 3059]]
Accuracy:
0.844
```

Fig. 4. Confusion Matrix and Accuracy Calculations – MNB

### c. Random Forest

. Using the Random Forest, the model gave the following,
Accuracy = 84.48%
False Positive Count = 551
False Negative Count = 613
The Confusion Matrix is shown in the below picture **Fig.5.** and Accuracy Calculations produced using the Random Forest,

```
Confusion matrix:
[[3126  613]
 [ 551 3210]]
Accuracy:
0.8448
```

Fig. 5. Confusion Matrix and Accuracy Calculations – RF

So, from the above results, we can say that the best results were obtained using Logistic Regression. amongst all the three models. Using random user inputs to demonstrate how these various models operated, with such simple and straightforward feedback all the models worked well and showed the correct output which is 1 indicating "Positive feedback". The results are shown in **Fig.6.** produced by our models when giving the feedback line = "It is an amazing insurance product" as an input in our model shows the output of the Sentiment score in order (top to bottom)– Logistic Regression, Random Forest, Multinomial Naive Bayes.

```
⇨  1
   1
   1
```

**Fig. 6. Sentiment Score**

With this twisted feedback, the Logistic Regression, as well as the Random Forest, generated the apt output which is 1 indicating "Positive feedback". Whereas the Multinomial Naïve Bayes did not make a satisfactory prediction of the sentiment by showing 0 indicating "Negative feedback". The results are shown in **Fig .7**. produced by our models when giving the feedback line = "This product is not that good" as an input in our model. Show the output of Logistic Regression, Random Forest, and Classifier Naive Bayes is the sentiment scores in order (top to bottom).

```
⇨  1
   1
   0
```

**Fig. 7. Sentiment Score**

Thus, being with the desired level of test precision the Logistic Regression Model proves to dominate in this dataset, the other two models are, Regarding Random Forest and the Multinomial Naïve Bayes both giving almost the same percentage of accuracy would be better off considered as at par to each other, as the results highly depend on the complexity of the feedback lines.

### 4.2    Recommendation Results

Let's first look at the shape of our dataset is shown in **Fig.8.** the first 5 tuples of our dataset – attributes (id, age, age lower range, age upper range, gender, income category, locality, id of the other product purchased, sentiment, review)

| | id | Age | Age_range_lower | Age_range_upper | Gender | Income Category | Locality | Other Product purcased by this customer | sentiment | review |
|---|---|---|---|---|---|---|---|---|---|---|
| 0 | 5814_8 | 47 | 45 | 50 | F | 1 | Outskirts | 6 | 1 | With all this stuff going down at the moment w... |
| 1 | 2381_9 | 47 | 45 | 50 | M | 1 | Outskirts | 5 | 1 | \The Classic War of the Worlds\" by Timothy Hi... |
| 2 | 7759_3 | 27 | 25 | 30 | F | 2 | Outskirts | 3 | 0 | The film starts with a manager (Nicholas Bell)... |
| 3 | 3630_4 | 34 | 30 | 35 | F | 2 | Outskirts | 2 | 0 | It must be assumed that those who praised this... |
| 4 | 9495_8 | 32 | 30 | 35 | M | 1 | City | 4 | 1 | Superbly trashy and wondrously unpretentious 8... |

**Fig. 8. Data Slice**

**Fig.9**. is shown the Customer attributes for whom the recommendation list of other insurance products is to be generated.

```
42
40
45
M
2
City
It is a great product
```

**Fig. 9. Customer attributes**

This feedback line was given as an input into our Sentiment Models and all the models performed well by indicating the feedback to be "positive feedback" – 1. Now, using this sentiment score and other attributes of our customer the following was generated as the product recommendation list for our customer as shown in **Fig.10.**,

```
Int64Index([4, 5, 3, 2, 6], dtype='int64')
```

**Fig. 10. Recommendation List**

This indicates that the best product to recommend is based on his liking/disliking of Product 1 and by comparing the customer attributes with all other customer's attributes. At the same time used on frequency analysis of each product purchased throughout the list of matching customers and hence displaying the result in an optimum manner.

So, Product 4 would be the first best recommendation for our customer, then goes product 5, 3, 2 and lastly product 6. The total frequency of each product purchased by all the customers in the matching list.is shown in **Fig.11.**

```
4    47
5    46
3    35
2    21
6     4
```

**Fig. 11. Frequencies**

As shown in **Fig.12.** the list of matching customers showcases that 153 out of 25000 of our existing customers best match our new customers.

| | id | Age | Age_range_lower | Age_range_upper | Gender | Income | Category | Locality | Other Product purcased by this customer | sentiment | |
|---|---|---|---|---|---|---|---|---|---|---|---|
| 77 | 6520_7 | 41 | 40 | 45 | M | 2 | | City | 3 | 1 | The original D |
| 105 | 4680_10 | 43 | 40 | 45 | M | 2 | | City | 5 | 1 | Terrfi |
| 269 | 11391_8 | 43 | 40 | 45 | M | 2 | | City | 3 | 1 | An American |
| 423 | 1757_8 | 44 | 40 | 45 | M | 2 | | City | 4 | 1 | The key to |
| 495 | 11456_10 | 42 | 40 | 45 | M | 2 | | City | 4 | 1 | Andy Gol |
| ... | ... | ... | ... | ... | ... | ... | | ... | ... | ... | ... |
| 24399 | 813_10 | 41 | 40 | 45 | M | 2 | | City | 5 | 1 | After "I |
| 24400 | 8464_10 | 43 | 40 | 45 | M | 2 | | City | 4 | 1 | Dear Readers |
| 24453 | 1658_10 | 42 | 40 | 45 | M | 2 | | City | 5 | 1 | If you've s |
| 24745 | 11032_7 | 42 | 40 | 45 | M | 2 | | City | 5 | 1 | Wrack |
| 24759 | 2907_7 | 43 | 40 | 45 | M | 2 | | City | 4 | 1 | In conce |

153 rows × 10 columns

**Fig. 12. Matches**

The overall result shows good reliability of our results. Out of 25000 customers, we filtered the best matching 153 customers for giving recommendations to our new customer based on the binary sentiment score generated using our sentiment model and then combining it further with the customer attributes to generate the apt recommendations.

## 5      Conclusion & Future Scope

The outcome from the modeling of sentiment to generating the list of top recommendable products was all convincing. The results of the sentiment modeling seemed very logical and sensible, plus, the recommendation list, as well as the matching list generated, could be of great help with relation to the sales and the analysis teams working in the insurance companies. Moreover, the recommendation results would also prove to be quite beneficial to the buyer. Despite all such benefits and results as there are flaws in every model as well as so does ours. This was a simple sentiment and recommendation modeling which required the data to be inappropriate format, the reviews to be collected related to the insurance product.

The sentiment modeling techniques can be enhanced using the BERT model for better analysis of the text and their sentiments and thus improving the sentiment modeling part. Then in the case of the recommendation part, there are many scopes for improvement like we may think to consider the correlation between different characteristics of the customers as well as consider that while matching for similarities between individuals. We could also think of generating a recommendation score by giving different weights to different attributes of the customer, maybe like giving more weights to age and income than gender as the formers play a greater influence on the insurance product preferences amongst individuals. Getting the chance to collect the data from an insurance giant might help us produce better recommendations as we probably would be having the actual recommendations made that resulted in more sales, hence, could test and find out whether our predicted recommendation lists are accurate or not.